\documentclass[%
superscriptaddress,
preprint,
showpacs,
amsmath,amssymb,
prc,
floatfix ]%
{revtex4-1}

\usepackage{color}
\usepackage{CJK}

\usepackage{graphicx}% Include figure files
\usepackage{dcolumn}% Align table columns on decimal point
\usepackage{bm}% bold math
\usepackage{hyperref}% add hypertext capabilities
\allowdisplaybreaks

\begin{document}

\begin{CJK*}{GB}{}
\title{Multidimensionally-constrained relativistic Hartree-Bogoliubov study of nuclear 
spontaneous fission}
\CJKfamily{gbsn}
\author{Jie Zhao (ÕÔ½Ü)}%
%\author{Jie Zhao}
\email{zhaojie@itp.ac.cn}
 \affiliation{Physics Department, Faculty of Science, University of Zagreb, Bijenicka 32,
              Zagreb 10000, Croatia}
% \affiliation{State Key Laboratory of Theoretical Physics,
%              Institute of Theoretical Physics, Chinese Academy of Sciences, Beijing 100190, China}
\CJKfamily{gbsn}
\author{Bing-Nan Lu (ÂÀ±þéª)}%
%\author{Bing-Nan Lu}
%\email{zhaojie@itp.ac.cn}
 \affiliation{Institut f\"ur Kernphysik, Institute for Advanced Simulation, 
	      and J\"ulich Center for Hadron Physics,
              Forschungszentrum J\"ulich, D-52425 J\"ulich, Germany}
% \affiliation{State Key Laboratory of Theoretical Physics,
%              Institute of Theoretical Physics, Chinese Academy of Sciences, Beijing 100190, China}
\author{Tamara Nik\v{s}i\'c}%
%\email{tniksic@phy.hr}
 \affiliation{Physics Department, Faculty of Science, University of Zagreb, Bijenicka 32,
              Zagreb 10000, Croatia}             
\author{Dario Vretenar}%
%\email{vretenar@phy.hr}
 \affiliation{Physics Department, Faculty of Science, University of Zagreb, Bijenicka 32,
              Zagreb 10000, Croatia}

\date{\today}

\begin{abstract}
\begin{description}
\item[Background]
Recent microscopic studies, based on the theoretical framework of 
nuclear energy density functionals, have analyzed dynamic (least action) and 
static (minimum energy) fission paths, and it has been shown that in addition to 
the important role played by nonaxial and/or octupole collective degrees of 
freedom, fission paths crucially depend on the approximations adopted in 
calculating the collective inertia.
\item[Purpose] To analyse effects of triaxial and octupole deformations, as 
well as approximations to the collective inertia, on the symmetric and asymmetric 
spontaneous fission dynamics, and compare with results of recent studies based 
on the self-consistent Hartree-Fock-Bogoliubov (HFB) method.
\item[Methods]
Deformation energy surfaces, collective potentials, and perturbative and nonperturbative 
cranking collective inertia tensors 
are calculated using the multidimensionally-constrained relativistic Hartree-Bogoliubov 
(MDC-RHB) model, with 
the energy density functionals PC-PK1 and DD-PC1.
Pairing correlations are treated in the Bogoliubov approximation using 
a separable pairing force of finite range. The least-action principle is employed to 
determine dynamic spontaneous fission paths.
\item[Results]
The dynamics of spontaneous fission of $^{264}$Fm and $^{250}$Fm is explored. 
The fission paths, action integrals and 
the corresponding half-lives predicted by the functionals PC-PK1 and DD-PC1 are compared and, 
in the case of $^{264}$Fm, discussed in relation with recent results obtained using the HFB model 
based on the Skyrme functional SkM$^*$ and a density dependent mixed pairing interaction. 
\item[Conclusions]
The inclusion of nonaxial quadrupole and octupole shape degrees of freedom is essential for a 
quantitative analysis of fission dynamics. The action integrals and, consequently, the half-lives crucially depend 
on the approximation used to calculate the effective collective inertia along the fission path. 
The perturbative cranking approach underestimates the effects of 
structural changes at the level crossings and the resulting collective inertia varies relatively smoothly 
in the $(\beta_{20},\beta_{22})$ and $(\beta_{20},\beta_{30})$ planes. In contrast, the 
nonperturbative 
collective mass is characterized by the occurrence of sharp peaks 
on the surface of collective coordinates, that can be related to single-particle level crossings 
near the Fermi surface. This enhances the effective inertia, 
increases the values of the action integral, and results in longer fission half-lives. 
\end{description}
\end{abstract}

\pacs{21.60.Jz, 24.75.+i, 25.85.Ca, 27.90.+b}
%21.60.Jz       Nuclear Density Functional Theory and extensions
%               (includes Hartree-Fock and random-phase approximations)
%24.75.+i       General properties of fission
%25.85.Ca 	Spontaneous fission
%27.90.+b       A >= 220

\maketitle

\end{CJK*}

\section{Introduction~\label{sec:Introduction}}

Spontaneous fission (SF) presents a complex quantum process of evolution of a nucleus from the initial ground state 
to the final state with two fragments, and includes tunneling through barrier(s) in a multidimensional collective 
space \cite{Krappe&Pomorski_2012}. A number of microscopic approaches, such as the time-dependent generator coordinate 
method (TDGCM)~\cite{Reinhard1987_RPP50-1,Goutte2005_PRC71-024316,Younes2012-LLNLTR_586678,
Berger1991_CPC63-365,Berger1984_NPA428-23}, the 
adiabatic time-dependent Hartree-Fock (ATDHF) or Hartree-Fock-Bogoliubov (ATDHFB) 
approximation~\cite{Ring1980,Dobaczewski1981_NPA369-123,Baranger1978_AoP114-123,Giannoni1980_PRC21-2060},
and  mean-field instantons~\cite{Negele1989_NPA502-371,Skalski2008_PRC77-064610},
have been developed to describe fission. However, the development and applications of these methods 
to realistic cases are far from complete.

In a semi-classical approximation the one-dimensional barrier penetration with the 
Wentzel-Kramers-Brillouin (WKB) approximation is usually used to evaluate the SF half-life.  
The fission path can be obtained either by minimizing the collective energy in the 
multidimensional space of coordinates which are used to describe the elongation of the nucleus (e.g. $\beta_{20}$),
or by minimizing the fission action integral in the collective space.
Static fission paths obtained by minimizing the collective energy computed with the
macroscopic-microscopic (MM) model \cite{Zhang2014_PRC90-054313,Bao2013_NPA906-1,Moeller1989_NPA492-349} 
and various self-consistent mean-field (SCMF) models~\cite{Warda2012_PRC86-014322,Warda2002_PRC66-014310,
Staszczak2013_PRC87-024320,Staszczak2009_PRC80-014309,Erler2012_PRC85-025802,
Schindzielorz2009_IJMPE18-773,Berger2001_NPA685-1,Giuliani2013_PRC88-054325,Rodriguez-Guzman2014_EPJA50-142}
have been used to calculate SF half-lives.

The concept of dynamic least-action fission path was introduced in 
Refs.~\cite{Brack1972_RMP44-320,Ledergerber1973_NPA207-1}, and subsequently 
effective methods to determine the dynamic path numerically in a multidimensional collective space 
were developed~\cite{Baran1978_PLB76-8,Baran1981_NPA361-83}.
Both the potential energy surface and inertia tensors are crucial in determining the dynamic fission path.
The ATDHFB method with the perturbative cranking approximation
(that is, neglecting the contribution from time-odd mean fields, and treating perturbatively the derivatives 
of the single-nucleon and pairing densities with respect to 
collective coordinates) has usually been used in SF fission life-time 
calculations~\cite{Brack1972_RMP44-320,Nilsson1969_NPA131-1,Girod1979_NPA330-40,Bes1961_NP28-42,
Sobiczewski1969_NPA131-67}.
The dynamic fission path, however, can differ significantly from the static one. 
For instance, it was shown that the triaxial quadrupole degree of freedom plays an important role 
around the inner and outer barriers along the static fission path for actinide nuclei 
(Ref.~\cite{Lu2014_PRC89-014323} and references therein).
Nevertheless, the effect of triaxiality on the dynamic fission path is negligible or small in the majority 
of cases~\cite{Gherghescu1999_NPA651-237,Smolanczuk1993_APPB24-685,Smolanczuk1995_PRC52-1871}.
The odd-multipole deformations $\beta_{30}$ and $\beta_{50}$ were also found to have a small 
effect on the dynamic fission path, whereas their inclusion lowers the static fission barrier 
considerably at large quadrupole deformations~\cite{Lojewski1999_NPA657-134}.
Studies of SF with the least-action principle have also shown that pairing vibrations have a   
pronounced effect on the fission probability~\cite{Moretto1974_PLB49-147,Staszczak1985_PLB161-227,
Staszczak1989_NPA504-589,Smolanczuk1993_APPB24-685,Giuliani2014_PRC90-054311,Sadhukhan2014_PRC90-061304,
Urin1966_NP75-101,Lazarev1987_PS35-255,Pomorski2007_IJMPE16-237}. 
Systematic investigations of SF half-lives of superheavy nuclei with the dynamic approach were performed 
based on the MM~\cite{Smolanczuk1995_PRC52-1871} and 
HFB models~\cite{Baran2005_PRC72-044310,Staszczak2013_PRC87-024320}. 

The nonperturbative cranking ATDHFB collective mass tensor for which the derivatives with respect to 
collective coordinates are calculated explicitly using numerical techniques, was recently used in 
illustrative calculations of one-dimensional quadrupole fission paths~\cite{Baran2011_PRC84-054321}. 
It was shown that the collective mass exhibits strong variations with the quadrupole collective coordinate, 
related to changes in the intrinsic shell structure. 
By using the nonperturbative cranking mass in SF dynamic studies~\cite{Sadhukhan2013_PRC88-064314}
marked triaxial effects were predicted along dynamic 
fission paths, consistent with those obtained in static calculations, whereas it was found that 
using the perturbative-cranking mass drives the system towards near-axial shapes. It was 
noted that the structural properties of the collective mass play an essential role in determining the 
SF dynamics.

Models based on the framework of relativistic nuclear energy density functionals have 
been successfully applied to the description of deformation energy landscapes and fission barriers of heavy 
and superheavy nuclei (Ref.~\cite{Lu2014_PRC89-014323} and references therein). 
By breaking both axial and reflection symmetries, 
the multidimensionally-constrained relativistic mean-field (MDC-RMF) 
and multidimensionally-constrained relativistic Hartree Bogoliubov (MDC-RHB) 
model have recently been developed and implemented in studies of deformation energy maps and fission 
barriers of actinide nuclei~\cite{Lu2012_PRC85-011301R,Lu2014_PRC89-014323,
Lu2012_EPJWoC38-05003,Lu2014_JPCS492-012014, Lu2014_PS89-054028,Lu2012_PhD,Zhao2015_PRC91-014321},
shapes of hypernuclei~\cite{Lu2011_PRC84-014328,Lu2014_PRC89-044307},
and nonaxial-octupole $Y_{32}$ correlations in $N = 150$ isotones~\cite{Zhao2012_PRC86-057304}.

In this work we explore the dynamics of SF using the MDC-RHB model.
The deformation energy surfaces of  $^{264}$Fm and $^{250}$Fm are computed by solving 
constrained RHB equations in a multidimensional collective coordinate space.
The collective inertia tensor is calculated using the self-consistent RHB solutions and  
applying the ATDHFB expressions in both the perturbative-cranking and nonperturbative-cranking approximations. 
The dynamic fission paths are determined by the least-action principle with perturbative-cranking
and nonperturbative-cranking inertias, and the corresponding SF half-lives are computed.
The article is organized as follows: the theoretical framework 
is introduced in Sec.~\ref{sec:model}, numerical details of the calculation, the results for 
the deformation energy landscapes, inertias, and minimum-action fission paths are 
discussed in Sec.~\ref{sec:results}, and a
summary and conclusions are included in Sec.~\ref{sec:summary}.

%%%%%%%%%%%%%%%%%%%%%%%%%%%%%%%%%%%%%%%%%%%%
\section{\label{sec:model}Theoretical framework}
%%%%%%%%%%%%%%%%%%%%%%%%%%%%%%%%%%%%%%%%%%%%

The tool of choice for theoretical studies of the structure of medium-heavy and heavy nuclei is
the framework of energy density functionals (EDFs)~\cite{Bender2003_RMP75-121,Vretenar2005_PR409-101}. 
Self-consistent mean-field models based on semi-empirical EDFs provide an accurate and reliable microscopic
description of nuclear structure phenomena over the entire nuclide chart. 
EDF-based structure models have also been developed that go beyond the static 
mean-field approximation, and include collective correlations 
related to the restoration of broken symmetries and to fluctuations of
collective variables. 
Relativistic mean-field (RMF) models present a particular implementation of the nuclear 
EDF framework. In the standard representation based on the Walecka model, the
atomic nucleus is described as a system of Dirac nucleons coupled to exchange mesons through
an effective Lagrangian. However, at the energy scale characteristic for nuclear binding
low-lying excitations, the meson exchange is just a convenient representation of the effective 
nuclear interaction, and can be replaced by the local contact interactions between nucleons. 
To describe nuclear properties at a quantitative level, higher order 
many-body effects have to be included through a medium dependence of the 
inter-nucleon interaction. This can be achieved either by including higher-order (nonlinear) terms in the 
Lagrangian, or by assuming an explicit density dependence for the vertex functions.
In the present study we employ two standard and representative point-coupling relativistic EDFs that have been extensively 
used in studies of a variety of nuclear properties. PC-PK1~\cite{Zhao2010_PRC82-054319} includes 
higher-order interaction terms in the nucleon self-energies, and DD-PC1~\cite{Niksic2008_PRC78-034318} 
with quadratic interaction terms but including explicit density-dependent vertex functions.  

For a quantitative description of open-shell nuclei it is necessary to consider also
pairing correlations. The relativistic Hartree-Bogoliubov (RHB) framework~\cite{Vretenar2005_PR409-101} provides 
a unified description of particle-hole ($ph$) and particle-particle correlations by combining
two average potentials: the self-consistent mean field $\Gamma$ that encloses all the long range $ph$
correlations, and a pairing field $\Delta$ which sums up the $pp$ correlations. Here we use a pairing
force separable in momentum pace in the $pp$ channel:
\begin{equation}
\langle k | V^{1S_0}| k^\prime \rangle = - Gp(k) p(k^\prime).
\end{equation}
A simple Gaussian ansatz $p(k)=e^{-a^2k^2}$ in momentum space is assumed and, 
when transformed from momentum to coordinate space,
the interaction takes the form:
\begin{equation}
V(\mathbf{r}_1,\mathbf{r}_2,\mathbf{r}_1^\prime,\mathbf{r}_2^\prime) = G_0 ~\delta(\mathbf{R}-
\mathbf{R}^\prime) P (\mathbf{r}) P(\mathbf{r}^\prime) \frac{1}{2} \left(1-P^\sigma\right),
\label{pairing}
\end{equation}
where $\mathbf{R} = (\mathbf{r}_1+\mathbf{r}_2)/2$ and $\mathbf{r}=\mathbf{r}_1- \mathbf{r}_2$
denote the center-of-mass and the relative coordinates, and $P(\mathbf{r})$ is the Fourier
transform of $p(k)$
\begin{equation}
P(\mathbf{r})=\frac{1}{\left(4\pi a^2\right)^{3/2}} e^{-\mathbf{r}^2/4a^2}.
\end{equation}
The two parameters $G_0$ and $a$ have been
adjusted to reproduce the density dependence of the pairing gap in nuclear matter at the
Fermi surface. The pairing gap calculated with the D1S parameterization of the Gogny force~\cite{Berger1991_CPC63-365} 
is reproduced using the interaction (\ref{pairing}) with the following values: 
$G_0=-738$ MeV fm$^{-3}$ and $a=0.644$ fm~\cite{Tian2009_PLB676-44}. 

The deformation energy landscape is obtained in a self-consistent mean-field 
calculation with constraints on mass 
multipole moments~\cite{Ring1980}. Here we 
use a modified linear-constraint method with the Routhian
defined as
\begin{equation}
E^\prime = E_{RHB} + \sum_{\lambda\mu}{\frac{1}{2}C_{\lambda \mu}Q_{\lambda \mu}}.
\end{equation}
In each iteration step the coefficients $C_{\lambda \mu}$ are modified: 
\begin{equation}
C_{\lambda \mu}^{(n+1)} = C_{\lambda \mu}^{(n)} + k_{\lambda \mu}
\left(  \beta^{(n)}_{\lambda \mu} -  \beta_{\lambda \mu}\right) ,
\end{equation}
where $\beta_{\lambda \mu}$ is the desired deformation, $k_{\lambda \mu}$ is a
constant, and $C_{\lambda \mu}^{(n)}$ denotes the value of the coefficient in the $n$-th 
iteration step.

To describe nuclei with general quadrupole and/or octupole shapes, the
Dirac-Hartree-Bogoliubov equations are solved by expanding
the nucleon spinors in the basis of a 3D harmonic oscillator in
Cartesian coordinates. Basis states satisfying $[n_z/Q_z+(2n_\rho+|m_l|)/Q_\rho] \leq N_f$
are included for the large component of the Dirac single-nucleon wave function,
where $Q_z=\max(1,b_z/b_0)$ and $Q_\rho=\max(1,b_\rho/b_0)$ are constants related to the 
oscillator lengths $b_0=1/\sqrt{M\omega_0}$, $b_z$, and $b_\rho$.
For the small component of the Dirac spinor $N_g=N_f+1$ major shells are included 
in order to avoid the occurrence of spurious states~\cite{Gambhir1990_APNY198-132}.
In the present study of transactinide
nuclei calculations have been performed in a basis with $N_f^{\rm max}=16$ shells.

The nuclear shape is parameterized by the deformation parameters
\begin{equation}
 \beta_{\lambda\mu} = {4\pi \over 3AR^\lambda} \langle Q_{\lambda\mu} \rangle,
\end{equation}
where $Q_{\lambda\mu} = r^\lambda Y_{\lambda \mu}$ is the mass multipole operator.
The shape is assumed to be invariant under the exchange of the $x$ and $y$ axes 
and all deformations $\beta_{\lambda\mu}$ with even $\mu$ can be included simultaneously.
For details of the MDC-RMF model we refer the reader to Ref.~\cite{Lu2014_PRC89-014323}.

We will explore the spontaneous fission (SF) process along a fission path $L$ that is
embedded in the multidimensional collective space. The path is defined by the
parameter $s$ with the inner ($s_{\rm in}$) and outer ($s_{\rm out}$) turning points.
The fission action integral reads
\begin{equation}
\label{eq:act_integration}
S(L) = \int_{s_{\rm in}}^{s_{\rm out}} {1\over\hbar} 
  \sqrt{ 2\mathcal{M}_{\rm eff}(s) \left[ V_{\rm eff}(s)-E_0 \right] } ds ,
\end{equation}
where $\mathcal{M}_{\rm eff}(s)$ and $V_{\rm eff}(s)$ are the effective 
collective inertia and potential along the fission path $L(s)$, respectively.
$E_0$ is the collective ground state energy, and the integration
limits correspond to the classical inner and outer turning points defined by: 
$V_{\rm eff}(s) = E_0$.
The fission path $L(s)$ is determined by minimizing the action
integral in Eq.~(\ref{eq:act_integration})~\cite{Brack1972_RMP44-320,Ledergerber1973_NPA207-1}.
The SF half-life is calculated as $T_{1/2}=\ln2/(nP)$, where $n$ is the
number of assaults on the fission barrier per unit 
time~\cite{Baran1978_PLB76-8,Baran1981_NPA361-83,
Sadhukhan2013_PRC88-064314,Sadhukhan2014_PRC90-061304}, 
and $P$ is the barrier penetration probability in the WKB approximation
\begin{equation}
P = {1 \over 1+\exp[2S(L)]}. 
\end{equation}

The essential ingredients in the calculation of the action integral -- expression (\ref{eq:act_integration}), are the effective collective
inertia and potential. The effective inertia is related to the multidimensional collective inertia tensor $\mathcal{M}$
\cite{Brack1972_RMP44-320,Baran1978_PLB76-8,Baran1981_NPA361-83,Sadhukhan2013_PRC88-064314,
Sadhukhan2014_PRC90-061304}
\begin{equation}
\mathcal{M}_{\rm eff}(s) = \sum_{ij} \mathcal{M}_{ij} {dq_i \over ds} {dq_j \over ds}\;,
\end{equation}
where $q_i(s)$ denotes the collective variable as function of the path's length. 

The collective inertia tensor is computed using the ATDHFB method \cite{Baran2011_PRC84-054321}.
In the nonperturbative cranking approximation the inertia tensor reads 
\begin{equation}
\label{eq:npmass}
\mathcal{M}_{ij}^{C} = {\hbar^2 \over 2 \dot{q}_i \dot{q}_j}
    \sum_{\alpha\beta} {F^{i*}_{\alpha\beta}F^{j}_{\alpha\beta} + F^{i}_{\alpha\beta}F^{j*}_{\alpha\beta}
    \over E_{\alpha} + E_{\beta}},
\end{equation}
where
\begin{equation}
\label{eq:fmatrix}
{F^{i} \over \dot{q}_{i}}  
  =  U^\dagger {\partial\rho \over \partial q_{i}} V^* 
    + U^\dagger {\partial\kappa \over \partial q_{i}} U^*
    - V^\dagger {\partial\rho^* \over \partial q_{i}} U^*
    - V^\dagger {\partial\kappa^* \over \partial q_{i}} V^*\;.
\end{equation}
$U$ and $V$ are the self-consistent Bogoliubov matrices, and $\rho$ and $\kappa$ are 
the corresponding particle and pairing density matrices, respectively.
The derivatives of the densities are calculated using the Lagrange three-point formula for 
unequally spaced points~\cite{Giannoni1980_PRC21-2076,Yuldashbaeva1999_PLB461-1}.
The formula Eq.~(\ref{eq:fmatrix}) can be further simplified by using a perturbative 
approach~\cite{Brack1972_RMP44-320,Nilsson1969_NPA131-1,Girod1979_NPA330-40,Bes1961_NP28-42,
Sobiczewski1969_NPA131-67}, with the resulting perturbative cranking inertia 
\begin{equation}
\label{eq:pmass}
\mathcal{M}^{Cp} = \hbar^2 {\it M}_{(1)}^{-1} {\it M}_{(3)} {\it M}_{(1)}^{-1}, 
\end{equation}
and with
\begin{equation}
\label{eq:mmatrix}
\left[ {\it M}_{(k)} \right]_{ij} = \sum_{\alpha\beta} 
    {\left\langle 0 \left| \hat{Q}_i \right| \alpha\beta \right\rangle
     \left\langle \alpha\beta \left| \hat{Q}_j \right| 0 \right\rangle
     \over (E_\alpha + E_\beta)^k}.
\end{equation}
$|\alpha\beta\rangle$ are two-quasiparticle wave functions.
Details of the derivation of the formulas for the inertia tensor can be found in 
Ref.~\cite{Baran2011_PRC84-054321}. 

The effective collective potential $V_{\rm eff}$ is obtained by subtracting the vibrational zero-point energy (ZPE) from the total RHB 
deformation energy. Following the prescription of 
Refs.~\cite{Staszczak2013_PRC87-024320,Baran2007_IJMPE16-443,Sadhukhan2013_PRC88-064314,Sadhukhan2014_PRC90-061304} 
the ZPE is computed using the Gaussian overlap approximation, 
\begin{equation}
\label{eq:zpe}
E_{\rm ZPE} = {1\over4} {\rm Tr} \left[ {\it M}_{(2)}^{-1} {\it M}_{(1)} \right],
\end{equation}
where the ${\it M}_{(k)}$ are given by Eq.~(\ref{eq:mmatrix}).
The microscopic self-consistent solutions of the constrained RHB equations, that is, the 
single-quasiparticle energies and wave functions on the entire energy surface as functions of the 
quadrupole deformations, provide the microscopic input for the calculation of the 
collective inertia and zero-point energy.

%----------------------------------------------------------------------------------------------------------------------
\section{\label{sec:results}Spontaneous fission of $^{\bf 264}{\rm \bf Fm}$ and $^{\bf 250}{\rm \bf Fm}$}
%----------------------------------------------------------------------------------------------------------------------

We analyse two illustrative examples: the symmetric spontaneous fission of 
$^{264}$Fm and the asymmetric SF of $^{250}$Fm.
Although in principle one could include arbitrary many collective coordinates in the description of the 
fission process, in practice available computational resources impose rather severe restrictions on 
the dimensionality of the collective space in self-consistent calculations.
The present study is restricted to a two-dimensional collective space defined by either
$(\beta_{20},\beta_{22})$ (quadrupole triaxial) or $(\beta_{20},\beta_{30})$ (quadrupole and 
octupole axial) collective coordinates.

Two relativistic NEDFs, PC-PK1 with 
nonlinear self-interaction terms~\cite{Zhao2010_PRC82-054319}, and DD-PC1
functional with density-dependent couplings~\cite{Niksic2008_PRC78-034318}, 
are used in the self-consistent RHB calculations of the deformation energy surfaces, 
collective inertia tensors and fission action integrals. We note that
the height of the fission barriers is rather sensitive to
the strength of the pairing interaction~\cite{Karatzikos2010_PLB689-72}. 
Thus, the particular choice of the pairing strength may considerably affect the fission dynamics.
As explained above, the parameters of the finite range separable pairing force 
were originally adjusted to reproduce the pairing gap at the 
Fermi surface in symmetric nuclear matter as calculated with the Gogny D1S force.
However, a number of studies based on the relativistic Hartree-Bogoliubov model
have shown that the pairing strength needs to be fine-tuned in some 
cases, especially for heavy nuclei~\cite{Wang2013_PRC87-054331,Afanasjev2013_PRC88-014320}.
In this study the pairing strengths are further adjusted to reproduce the available empirical
pairing gaps in Fm isotopes.
%
%experimental odd-even 
%mass difference of $^{252}$Fm isotope:
% \begin{align}
% \Delta_n(Z,N)&=\frac{1}{2}\left[E(Z,N+1)-2E(Z,N)+E(Z,N-1)\right],\nonumber \\
%  \Delta_p(Z,N)&=\frac{1}{2}\left[E(Z+1,N)-2E(Z,N)+E(Z-1,N)\right] \nonumber.
% \end{align}
 The resulting values with respect to the original 
pairing strength adjusted in nuclear matter ($G_0=-738$ MeV fm$^{-3}$)are: 
$G_n/G_0=1.06$, $G_p/G_0=1.04$ for PC-PK1, 
and $G_n/G_0=1.11$, $G_p/G_0=1.08$ for DD-PC1.
As in Refs.~\cite{Sadhukhan2013_PRC88-064314,Sadhukhan2014_PRC90-061304}, 
we choose $E_0 =1$ MeV in Eq.~(\ref{eq:act_integration}) for the value of the collective ground state energy. 
Although arbitrary, this choice enables a direct comparison of our results with those reported in previous studies. 
For the vibrational frequency $\hbar\omega_0=1$ MeV the number of assaults on the fission barrier per unit 
is $10^{20.38}$ s$^{-1}$~\cite{Baran2005_PRC72-044310}.

%-----------------------------------------
\subsection{\label{subsec:Fm264}Symmetric fission of $^{\bf 264}{\rm \bf Fm}$}
%-----------------------------------------
%----
\begin{figure}
 \includegraphics[width=0.6\textwidth]{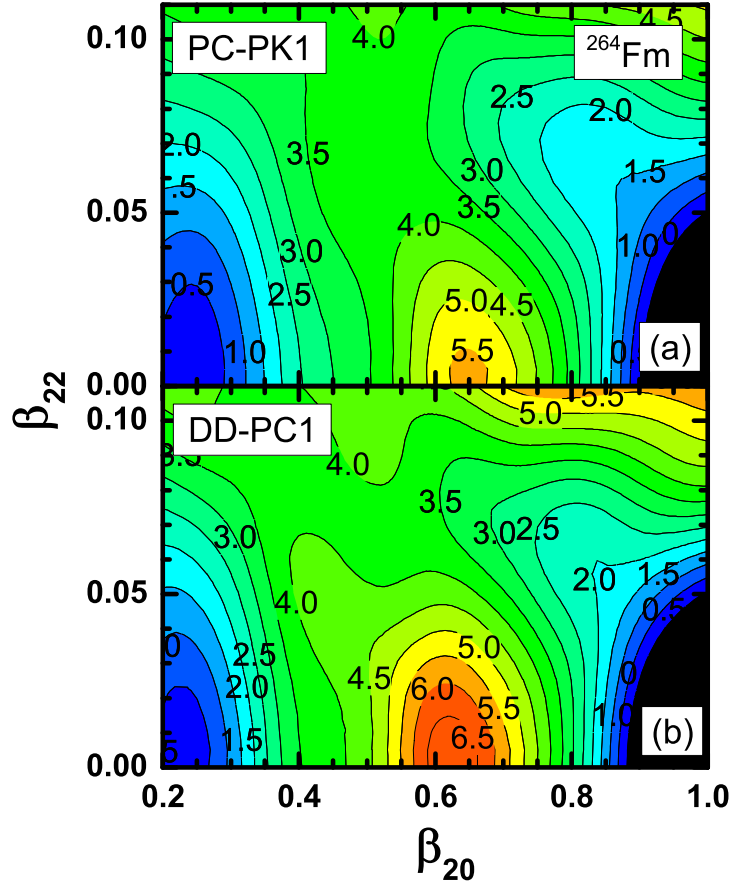}
\caption{(Color online)~\label{fig:Fm264_etot}%
RHB self-consistent triaxial quadrupole constrained energy surfaces of $^{264}$Fm in
the $(\beta_{20},\beta_{22})$ plane. In each panel energies are normalized with respect to the binding energy
of the equilibrium minimum, and contours join points on the surface with the same energy (in MeV). 
The energy surfaces in the upper (lower) panel are calculated with the density functionals 
PC-PK1~\cite{Zhao2010_PRC82-054319} (DD-PC1 \cite{Niksic2008_PRC78-034318}), 
and the pairing interaction Eq.~(\ref{pairing}).
}
\end{figure}
%-----
Previous theoretical studies of 
$^{264}$Fm~\cite{Staszczak2009_PRC80-014309,Staszczak2011_IJMPE20-552} have shown that 
one can expect this nucleus to undergo symmetric spontaneous fission and, therefore, we do not 
consider reflection-asymmetric degrees of freedom and perform the analysis in 
the collective space $(\beta_{20},\beta_{22})$. 
Figure \ref{fig:Fm264_etot} displays the self-consistent triaxial quadrupole deformation energy
surfaces of $^{264}$Fm in the $(\beta_{20},\beta_{22})$ plane. The energy surfaces in the upper 
(lower) panel are calculated with the density functionals 
PC-PK1 (DD-PC1), and the pairing interaction Eq.~(\ref{pairing})
The functional  PC-PK1 predicts an axially symmetric equilibrium (ground) state with moderate elongation
($\beta_{20} \approx 0.2$). The axially symmetric barrier at $\beta_{20}\approx 0.6$ is bypassed
through the triaxial region, thus lowering the height of the barrier by $\approx 2$ MeV.
With DD-PC1 a similar energy surface is obtained, however, with a more pronounced influence 
of the triaxial degree of freedom on the height of the barrier.

%----
\begin{figure}
 \includegraphics[width=0.6\textwidth]{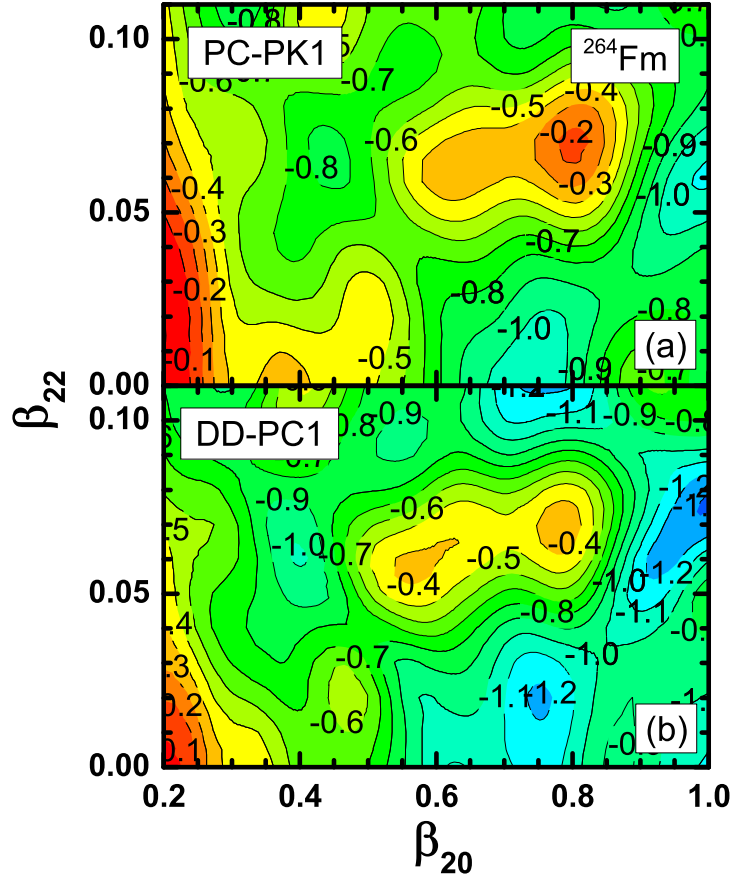}
\caption{(Color online)~\label{fig:Fm264_zpe}%
The vibrational zero-point energies $E_{\rm ZPE}$ Eq.~(\ref{eq:zpe}) of $^{264}$Fm in the $(\beta_{20},\beta_{22})$ plane. 
Energy surfaces obtained with the PC-PK1 and DD-PC1 functionals are compared
in the upper and lower panel, respectively. In each panel energies are normalized with respect to the 
the equilibrium minimum, and contours join points on the surface with the same energy (in MeV).}
\end{figure}
%----

The collective potential is obtained by subtracting the vibrational ZPE ($E_{\rm ZPE}$)
from the total binding energy surface. In Fig.~\ref{fig:Fm264_zpe} we plot the vibrational ZPE
Eq.~(\ref{eq:zpe}), normalized with respect to the mean-field ground 
state. The two functionals lead to rather similar results, and the deformation dependence of the
ZPEs is comparable to the results obtained in Ref.~\cite{Sadhukhan2013_PRC88-064314} 
using the Skyrme energy density functional SkM$^*$ and a density dependent mixed pairing 
interaction.

%----
\begin{figure}
 \includegraphics[width=0.6\textwidth]{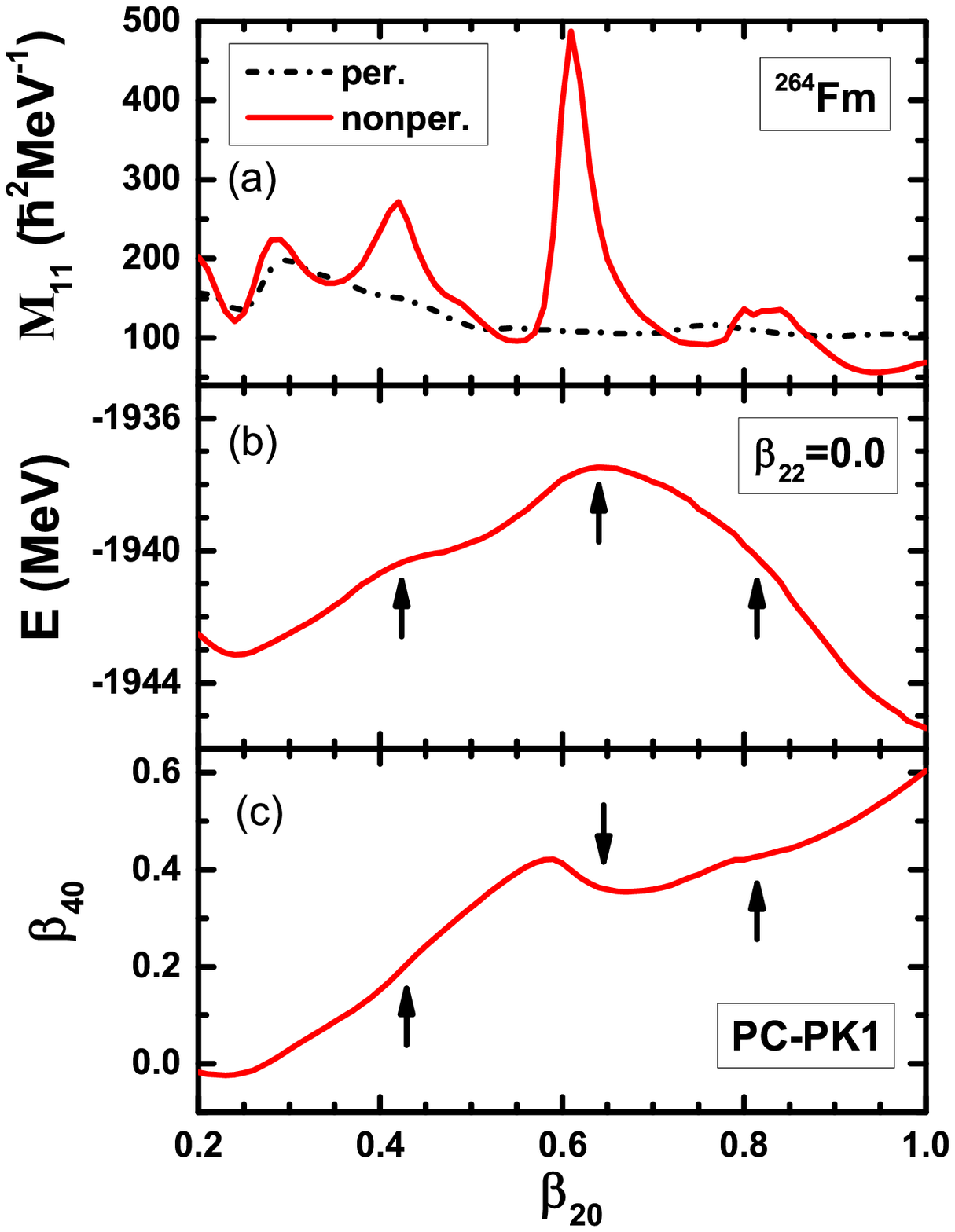}
\caption{(Color online)~\label{fig:tt_mass}%
The $\mathcal{M}_{11}$ component of the inertia tensor (top panel),  the binding energy
(middle panel), and the self-consistent  deformation parameter $\beta_{40}$ (bottom panel) 
of $^{264}$Fm as functions of the deformation $\beta_{20}$. Axial symmetry $\beta_{22}=0$ 
is imposed, and the functional PC-PK1 is used in the RHB calculation. 
}
\end{figure}
%----

To calculate the fission action integral one has to compute the collective inertia
tensor $\mathcal{M}_{ij}$. Although perturbative cranking mass parameters 
have been used in numerous studies, the importance of the exact treatment of 
derivatives of single-particle and pairing densities in the ATDHFB expressions 
for the mass parameters has recently been emphasized~\cite{Sadhukhan2013_PRC88-064314}. 
For the two-dimensional space of collective deformation coordinates 
three independent components $\mathcal{M}_{11}$, $\mathcal{M}_{12}$ and
$\mathcal{M}_{22}$ determine the inertia tensor and, in this case, the indices $1$ and $2$ 
refer to the $\beta_{20}$ and $\beta_{22}$ degrees of freedom, respectively.
The difference between the perturbative and nonperturbative cranking approximations 
is clearly seen in the top panel of Fig.~\ref{fig:tt_mass},  
where we plot the $\mathcal{M}_{11}$ component of the collective inertia tensor as a function 
of $\beta_{20}$ for axial symmetry ($\beta_{22}=0$).
The solid (red) curve denotes the nonperturbative cranking mass parameter, whereas the 
dot-dashed (black) curve corresponds to the perturbative cranking mass parameter. 
$\mathcal{M}_{11}^{Cp}$ displays a smooth dependence on the deformation
parameter $\beta_{20}$ and, although one notices some fluctuations, their magnitude is small.
The deformation dependence of the nonperturbative cranking mass parameter $\mathcal{M}_{11}^{C}$
follows the perturbative result $\mathcal{M}_{11}^{Cp}$, however, several sharp peaks occur  
at deformations $\beta_{20}\approx 0.4$, $\beta_{20}\approx 0.6$ and $\beta_{20}\approx 0.8$.
To understand better these results, in Fig.~\ref{fig:tt_mass} we also
plot the binding energy (middle panel) and the self-consistent value of the
$\beta_{40}$ deformation parameter. We notice that the most pronounced peak,
located at $\beta_{20}\approx 0.6$,  actually corresponds to the position of the
fission barrier. In general, the occurrence of sharp peaks in the collective mass is 
related to single-particle level crossings near the Fermi surface, that is, to abrupt changes 
of occupied single-particle configurations in a specific 
nucleus~\cite{Baran2011_PRC84-054321,Sadhukhan2013_PRC88-064314}.
Such pronounced structural rearrangements lead to strong variations in the 
derivatives of densities in Eq.~(\ref{eq:fmatrix}), and consequently sharp peaks develop in
the nonperturbative cranking collective inertia. At these specific 
deformations the value of the nonperturbative collective inertia can be several times larger
than the corresponding perturbative inertia, and this shows
that the effects of level crossing are not properly taken into account in the
perturbative cranking approach.

%----
\begin{figure}
 \includegraphics[width=0.6\textwidth]{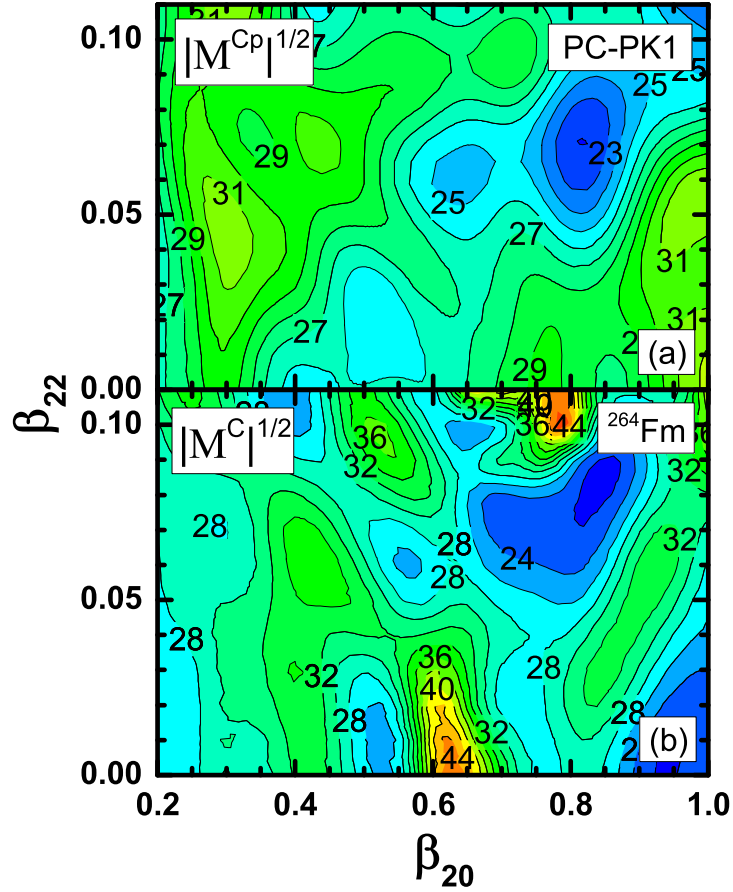}
\caption{(Color online)~\label{fig:Fm264_PCPK1_mass}%
Square-root determinants of the perturbative-cranking inertia tensor $|\mathcal{M}^{Cp}|^{1/2}$ (upper panel),
and nonperturbative-cranking inertia tensor $|\mathcal{M}^{C}|^{1/2}$ (lower panel)
(in $10 \times \hbar^2$MeV$^{-1}$), of $^{264}$Fm in the $(\beta_{20},\beta_{22})$ plane. 
The functional PC-PK1 is used in the RHB calculation.
}
\end{figure}
%----

\begin{figure}
 \includegraphics[width=0.6\textwidth]{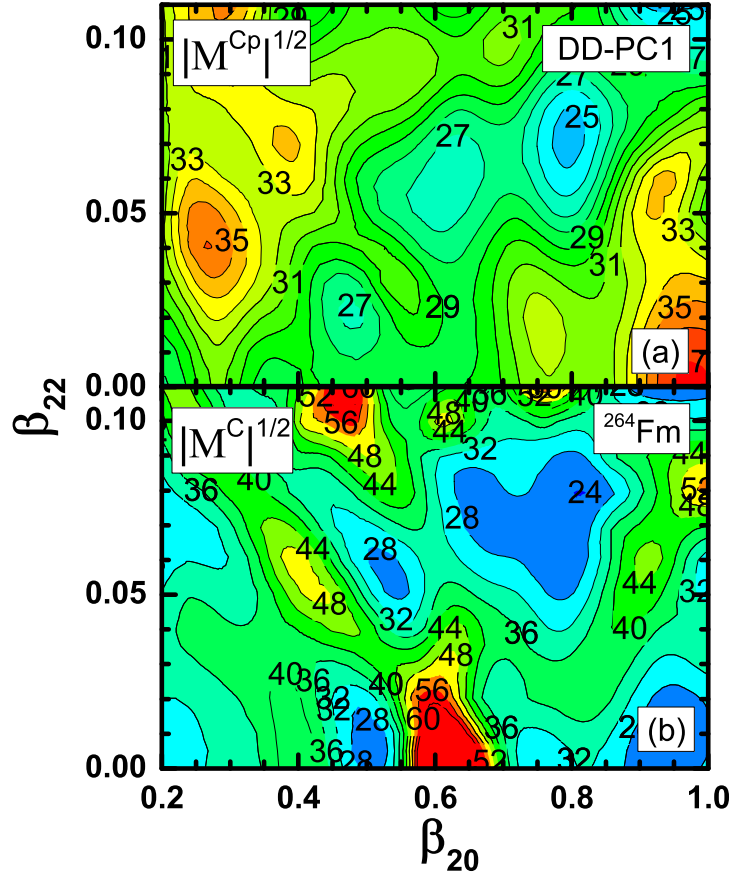}
\caption{(Color online)~\label{fig:Fm264_DDPC1_mass}%
Same as described in the caption to Fig.~\ref{fig:Fm264_PCPK1_mass} but for the functional DD-PC1.
}
\end{figure}

The collective inertia tensors can be visualized by plotting the 
square-root determinant
\begin{equation}
|\mathcal{M}|^{1/2}=(\mathcal{M}_{11} \mathcal{M}_{22} - \mathcal{M}_{12}^2)^{1/2},
\end{equation}
invariant with respect to rotations in the two-dimensional collective 
space~\cite{Sadhukhan2013_PRC88-064314}. In Figs.~\ref{fig:Fm264_PCPK1_mass} 
and~\ref{fig:Fm264_DDPC1_mass} we compare results obtained in a triaxial 
calculation with the perturbative and
nonperturbative approaches, and using the functionals PC-PK1 and the DD-PC1.
Although both approaches lead to rather complex 
topographies of $|\mathcal{M}|^{1/2}$ in the $(\beta_{20},\beta_{22})$ plane, we note 
more pronounced variations  for the nonperturbative approach.
In particular, the nonperturbative calculation results 
in very large values of $|\mathcal{M}|^{1/2}$ in the  region of the axial fission 
barrier, consistent with the
behaviour of the  component $\mathcal{M}_{11}$ for the axial case (cf. Fig.~\ref{fig:tt_mass}). 

%----
%----

The minimum action path is determined using two different numerical minimization techniques:
the dynamic-programming method (DPM)~\cite{Baran1981_NPA361-83}, and
the Ritz method (RM)~\cite{Baran1978_PLB76-8}. The DPM is implemented by discretizing
the energy surface in the $(\beta_{20},\beta_{22})$ plane with an equidistant two-dimensional mesh.
After considering all possible combinations of mesh points, the fission path is
constructed by connecting those points that minimize the action integral.
The RM, on the other hand, is implemented by expressing the trial path as a Fourier series
of collective coordinates. The coefficients of this series are determined by minimizing
the action integral. We note that for both methods we have considered several 
possible values for the turning points $s_{\rm in}$ and $s_{\rm out}$ to make certain that
the minimum action path is chosen. Details about the implementation of both the DPM and the RM 
are included in Appendix \ref{Appendix-DPM-RM}.

%----
\begin{figure}
 \includegraphics[scale=0.6]{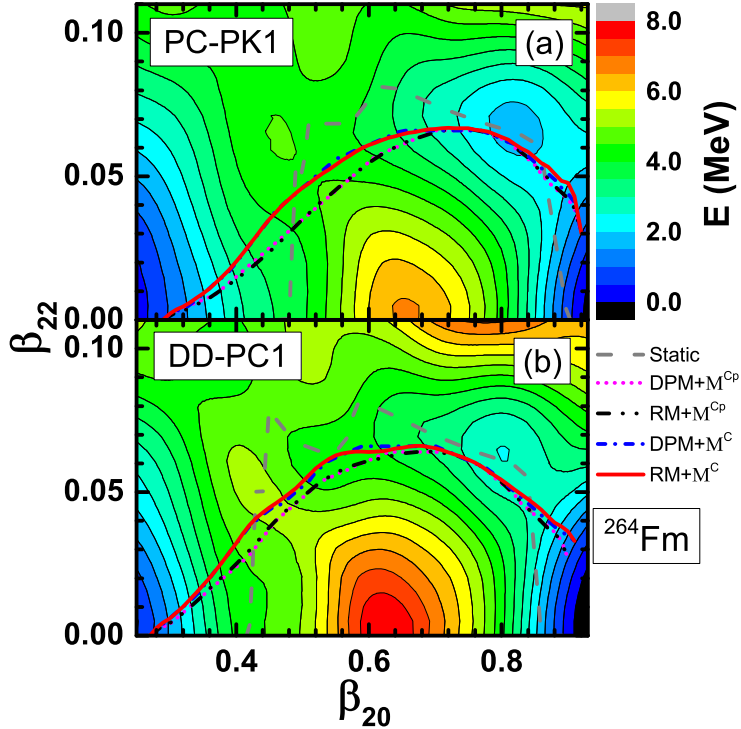}
\caption{(Color online)~\label{fig:Fm264_path}%
Dynamic paths for spontaneous fission of $^{264}$Fm in the $(\beta_{20},\beta_{22})$ 
plane, calculated 
with the functionals PC-PK1 (upper panel) and DD-PC1 (lower panel). 
The nonperturbative and perturbative cranking inertia are used, together with 
the DMP and RM techniques for the minimization of the collective 
action. The dotted and dash-dot-dot curves denote paths calculated with 
the perturbative-cranking inertia tensors using the DMP and RM, respectively, while the 
corresponding paths obtained with the nonperturbative-cranking inertia are plotted 
with the dash-dotted (solid) curves. The static path (dashed curve) is also shown for comparison.
}
\end{figure}
%----

The spontaneous fission paths on the triaxial deformation energy surface of $^{264}$Fm are shown
in Fig.~\ref{fig:Fm264_path}. Four different paths are included in the figure:
DPM + $\mathcal{M}^{Cp}$ path (dotted line), RM+$\mathcal{M}^{Cp}$ (dash-dot-dot),
DPM + $\mathcal{M}^{C}$ path (dash-dotted) and RM+$\mathcal{M}^{C}$ (solid).
The static path, determined by following the points of minimum energy between 
the turning points, is also shown for comparison (long dashed). It is interesting to note that, 
although they correspond to completely different effective interactions and were adjusted to data 
following different procedures, both PC-PK1 and DD-PC1 predict very similar paths. 
Similar SF paths are also obtained using the
perturbative and nonperturbative cranking inertia parameters. The paths 
detour the axial barrier through the triaxial region, although the excursion to the triaxial 
region is more pronounced for the nonperturbative cranking inertia. 
The present result differs somewhat from those obtained using the 
macroscopic-microscopic
approach~\cite{Gherghescu1999_NPA651-237,
Smolanczuk1993_APPB24-685,
Smolanczuk1995_PRC52-1871},
and the nonrelativistic HFB model~\cite{Sadhukhan2013_PRC88-064314,
Warda2012_PRC86-014322}, where
for the perturbative cranking inertia the nucleus $^{264}$Fm chooses an almost axially symmetric path towards fission. 
%--------
\begin{table}
%\vspace{-3mm}
%\footnotesize
%\begin{ruledtabular}
\begin{tabular}{llcr}
%\begin{tabular*}{170mm}{@{\extracolsep{\fill}}ccccccccc}
\hline \hline
 EDF & Path                           & S(L)     & $\log_{10}(T_{1/2}/{\rm yr})$ \\ \hline 
 PC-PK1    & Static + $\mathcal{M}^{Cp}$  &  $18.52$  & $-11.96$ \\
                 & Static + $\mathcal{M}^{C}$   & $19.69$     & $-10.94$ \\
                 & DPM+$\mathcal{M}^{Cp}$         & $14.52$  & $-15.43$                \\
                 & RM+$\mathcal{M}^{Cp}$          & $14.49$  & $-15.45$                \\
                & DPM+$\mathcal{M}^{C}$          & $15.53$  & $-14.55$                \\
                & RM+$\mathcal{M}^{C}$           & $15.48$  & $-14.59$                \\ \hline
DD-PC1   & Static + $\mathcal{M}^{Cp}$  &  $23.71$  & $-7.44$ \\
                 & Static + $\mathcal{M}^{C}$   & $27.07$     & $-4.53$ \\  
               & DPM+$\mathcal{M}^{Cp}$         & $17.84$  & $-12.54$                \\
               & RM+$\mathcal{M}^{Cp}$          & $17.81$  & $-12.57$                \\
               & DPM+$\mathcal{M}^{C}$          & $19.74$  & $-10.89$                \\
               & RM+$\mathcal{M}^{C}$           & $19.71$  & $-10.91$                \\ \hline
 SkM*~\cite{Sadhukhan2013_PRC88-064314}    
 & Static +$\mathcal{M}^{Cp}$  &  $20.8$  & $-10.0$ \\
 & Static + $\mathcal{M}^{C}$   & $23.4$     & $-7.7$ \\
 & DPM+ $\mathcal{M}^{Cp}$  
 &  $16.8$  & $-13.4$  \\
             & RM+$\mathcal{M}^{Cp}$          & $16.8$  &  $-13.4$  \\
             & DPM+$\mathcal{M}^{C}$          &  $19.1$  & $-11.4$ \\
             & RM+$\mathcal{M}^{C}$             &   $18.9$  & $-11.6$ \\ \hline \hline
\end{tabular}
%\end{ruledtabular}
\caption{\label{tab:action-Fm264} %
Values for the action integral and SF half-lives of $^{264}$Fm that correspond to the 
paths displayed in Fig.~\ref{fig:Fm264_path}. The results 
obtained in the present analysis (PC-PK1 and DD-PC1) are compared 
with those from Ref.~\cite{Sadhukhan2013_PRC88-064314}.}
\end{table}
%--------
The resulting values of the action integral and the fission half-lives are summarized in
Table~\ref{tab:action-Fm264}. Although very similar paths are obtained with 
PC-PK1 and DD-PC1, the corresponding values of the action integral differ by more than
 20\%, which leads to orders of magnitude difference in the calculated fission half-lives.
The difference between the perturbative and nonperturbative ATDHFB approximations 
for the collective inertia parameter is consistent for both functionals. We also note that
both minimization techniques produce virtually identical results for the action integral, 
and this provides a reliable test for the numerical accuracy and stability of the 
present calculation. Finally, it appears that the results obtained with the functional 
DD-PC1 are somewhat closer to those of Ref.~\cite{Sadhukhan2013_PRC88-064314}, 
calculated with the Skyrme functional SkM*, and very similar in the dynamical 
calculation with the nonperturbative cranking collective inertia.
We find that the triaxial degree of freedom always plays an important role in 
SF dynamics, independent of the approximation used to compute the inertia tensor.
However, the calculated half-lives are sensitive to the collective inertia.
The nonperturbative cranking mass predicts larger values of the 
fission action integral $S(L)$ and, therefore, longer half-lives.

%-----------------------------------------
\subsection{\label{subsec:Fm250}Asymmetric fission of $^{\bf 250}{\rm \bf Fm}$}
%-----------------------------------------

In the next example we explore the influence of the reflection-asymmetric
degree of freedom on the spontaneous fission process and study the  
asymmetric spontaneous fission $^{250}$Fm \cite{Staszczak2011_IJMPE20-552}.
Since the complete calculation in the three-dimensional collective space ($\beta_{20}$, $\beta_{22}$
and $\beta_{30}$) is computationally too demanding, we simplify the problem by determining the spontaneous 
fission dynamic path in two intervals: i) the 
path that connects the mean-field ground state and the isomeric state is calculated in the
$(\beta_{20},\beta_{22})$ plane, and ii) the path between the isomeric
state and the outer turning point is determined in the  $(\beta_{20},\beta_{30})$ plane.
The optimal path is obtained by combining the paths in the $(\beta_{20},\beta_{22})$ and
$(\beta_{20},\beta_{30})$ plane with the isomeric state as the matching point.
%----
\begin{figure}
 \includegraphics[width=0.6\textwidth]{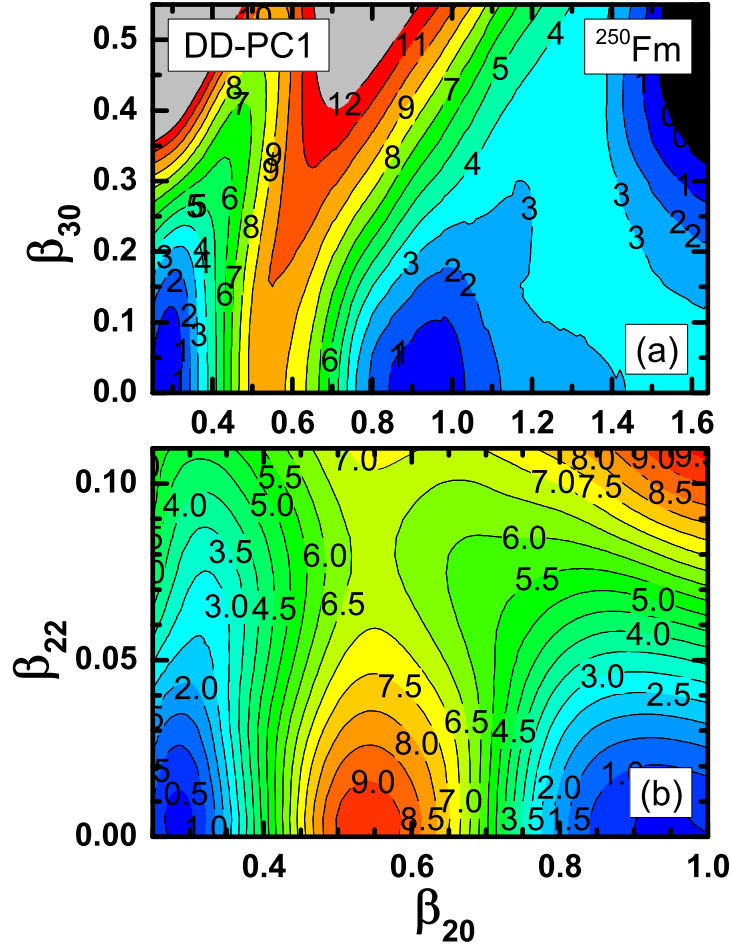}
\caption{(Color online)~\label{fig:Fm250_DDPC1_etot}%
RHB (DD-PC1 plus separable pairing) self-consistent constrained energy surfaces of $^{250}$Fm in 
in the $(\beta_{20},\beta_{30})$ 
(upper panel) and $(\beta_{20},\beta_{22})$ (lower panel) planes.
In each panel energies are normalized with respect to the binding energy
of the equilibrium minimum, and contours join points on the surface with the same energy (in MeV).
}
\end{figure}
%-----

\begin{figure}
 \includegraphics[width=0.6\textwidth]{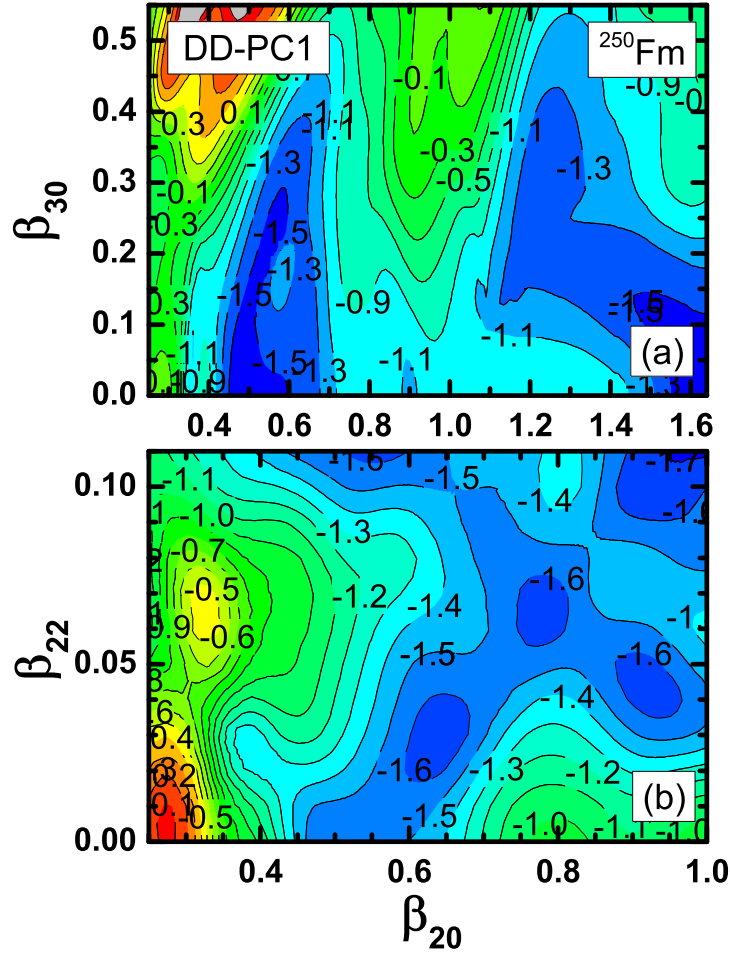}
\caption{(Color online)~\label{fig:Fm250_DDPC1_zpe}%
The vibrational zero-point energies $E_{\rm ZPE}$ Eq.~(\ref{eq:zpe}) of $^{250}$Fm in the $(\beta_{20},\beta_{30})$ 
plane (upper panel), and the $(\beta_{20},\beta_{22})$ plane (lower panel). 
Energies are normalized with respect to the 
the equilibrium minimum, and contours join points with the same energy (in MeV).
The functional DD-PC1 is used in the RHB calculation.
}
\end{figure}

In the upper panel of Fig.~\ref{fig:Fm250_DDPC1_etot} we display the RHB (DD-PC1 plus separable pairing)
deformation energy energy surface of $^{250}$Fm in the 
$(\beta_{20},\beta_{30})$  plane. The mean-field equilibrium (ground) state is predicted at moderate
quadrupole deformation $\beta_{20}\approx 0.3$, and the isomeric minimum at 
$\beta_{20}\approx 0.95$. We note that through the entire region
of quadrupole deformations $\beta_{20} \le 1.4$, the nucleus remains reflection symmetric,
that is, octupole degrees of freedom need not be included for this range of quadrupole deformations.
The region around the inner fission barrier is further analyzed in the lower panel of Fig.~\ref{fig:Fm250_DDPC1_etot}, 
where we plot the energy surface of $^{250}$Fm in the $(\beta_{20},\beta_{22})$  plane. Note the 
different horizontal scales in the two panels. 
The inclusion of the triaxial degree of freedom lowers the barrier by $\approx$ 2 MeV, 
and this effect is similar in magnitude to the case of $^{264}$Fm analyzed in the previous section.
Since triaxial shapes have the largest effect in the region of the first fission barrier, and 
reflection-asymmetric degrees of freedom are important for large quadrupole deformations, 
dividing the fission path into two segments provides a reasonable approximation for the complex 
multidimensional fission process. The vibrational zero-point energies of 
$^{250}$Fm isotope in the $(\beta_{20},\beta_{22})$ plane (lower panel) and
the  $(\beta_{20},\beta_{30})$ plane (upper panel) are shown in Fig.~\ref{fig:Fm250_DDPC1_zpe}.
For the whole deformation range considered in this figure the variation of $E_{\rm ZPE}$ is 
approximately 2 MeV, and very similar results are obtained with the functional PC-PK1. Note, 
however, the difference of the ZPE in the lower panel with respect to the quadrupole 
zero-point energy of $^{264}$Fm shown in Fig.~\ref{fig:Fm264_zpe}.

%--------
%--------

%----
\begin{figure}
 \includegraphics[width=0.6\textwidth]{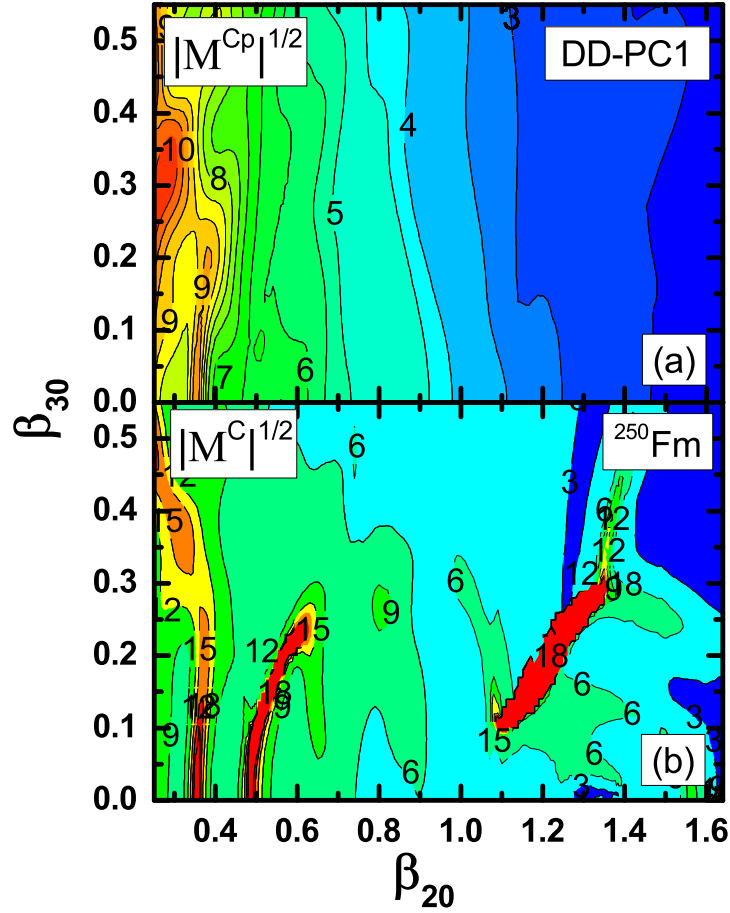}
\caption{(Color online)~\label{fig:Fm250_FT_DDPC1_mass}%
Square-root determinants of the perturbative cranking inertia tensor $|\mathcal{M}^{Cp}|^{1/2}$ (upper panel),
and nonperturbative cranking inertia tensor $|\mathcal{M}^{C}|^{1/2}$ (lower panel)
(in $10 \times \hbar^2$MeV$^{-1}$), of $^{250}$Fm in the $(\beta_{20},\beta_{30})$ plane. 
The calculation corresponds to axially symmetric but reflection asymmetric shapes.
}
\end{figure}
%----

\begin{figure}
 \includegraphics[width=0.6\textwidth]{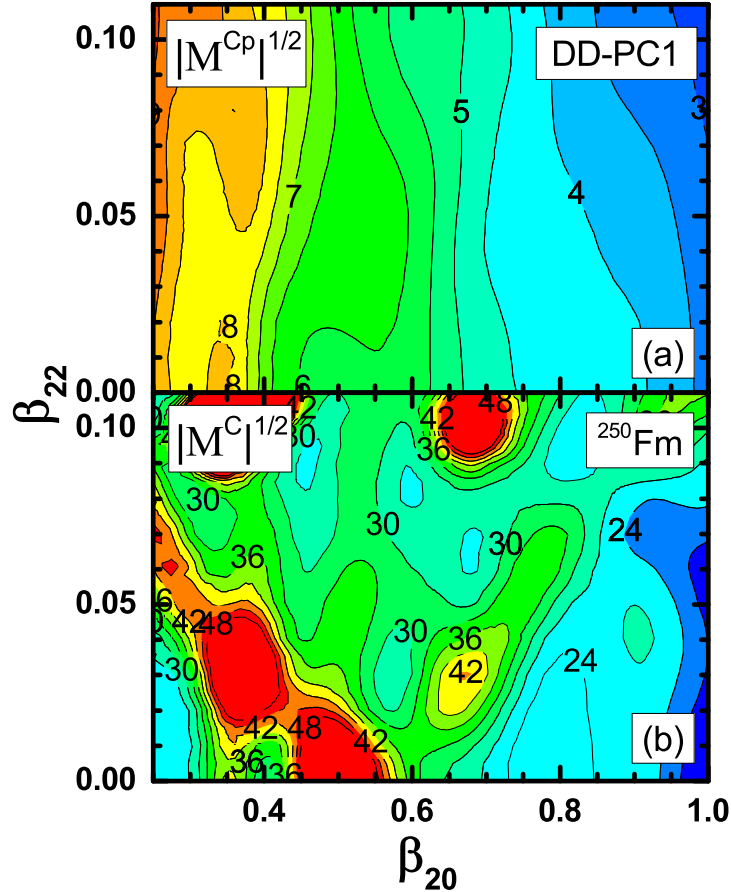}
\caption{(Color online)~\label{fig:Fm250_TF_DDPC1_mass}%
Same as described in the caption to Fig.~\ref{fig:Fm250_FT_DDPC1_mass} but for the 
inertia tensor in the $(\beta_{20},\beta_{22})$ plane. In this case the shape is allowed
to be triaxial but reflection symmetry is assumed.}
\end{figure}

The deformation dependence of the collective inertia tensor is illustrated in Figs.~\ref{fig:Fm250_FT_DDPC1_mass} 
and \ref{fig:Fm250_TF_DDPC1_mass}, where we plot the square-root determinant
$|\mathcal{M}|^{1/2} = \left( \mathcal{M}_{11} \mathcal{M}_{22} - \mathcal{M}_{12}^2\right)^{1/2}$ 
in the $(\beta_{20},\beta_{30})$ and $(\beta_{20},\beta_{22})$ planes, respectively. 
The perturbative cranking inertias $|\mathcal{M}^{Cp}|^{1/2}$ are shown in the upper panels, and 
the lower panels display the square-root determinants $|\mathcal{M}^{C}|^{1/2}$ of the nonperturbative 
cranking inertia tensor. The calculation of Fig.~\ref{fig:Fm250_FT_DDPC1_mass} corresponds to axially symmetric but 
reflection asymmetric shapes, that is, the indices 1 and 2 denote the $\beta_{20}$ 
and $\beta_{30}$ collective degrees of freedom, respectively. Figure \ref{fig:Fm250_TF_DDPC1_mass} shows the 
deformation dependence of the square-root determinants of collective inertia when the shape is allowed
to be triaxial but reflection symmetry is assumed. In this case the indices $1$ and $2$ denote the 
coordinates $\beta_{20}$ and $\beta_{22}$, respectively. The overall deformation dependence 
of $|\mathcal{M}^{Cp}|^{1/2}$ and $|\mathcal{M}^{C}|^{1/2}$ in Fig.~\ref{fig:Fm250_FT_DDPC1_mass} 
is similar but the nonperturbative cranking mass parameter displays several sharp peaks due to 
the crossing of single-particle levels around the Fermi surface (see the discussion in the previous section).
The picture is markedly different in the triaxial but reflection symmetric case illustrated in Fig.~\ref{fig:Fm250_TF_DDPC1_mass},
where the square-root determinant of the perturbative cranking collective inertia exhibits a smooth dependence in both 
$\beta_{20}$ and $\beta_{22}$ directions, whereas the nonperturbative
cranking inertia displays rapid fluctuations with very pronounced peak values caused by the level crossing effect.
Similar results are  also obtained with the functional PC-PK1, not shown here. 
%----
%----

\begin{figure}
 \includegraphics[width=0.6\textwidth]{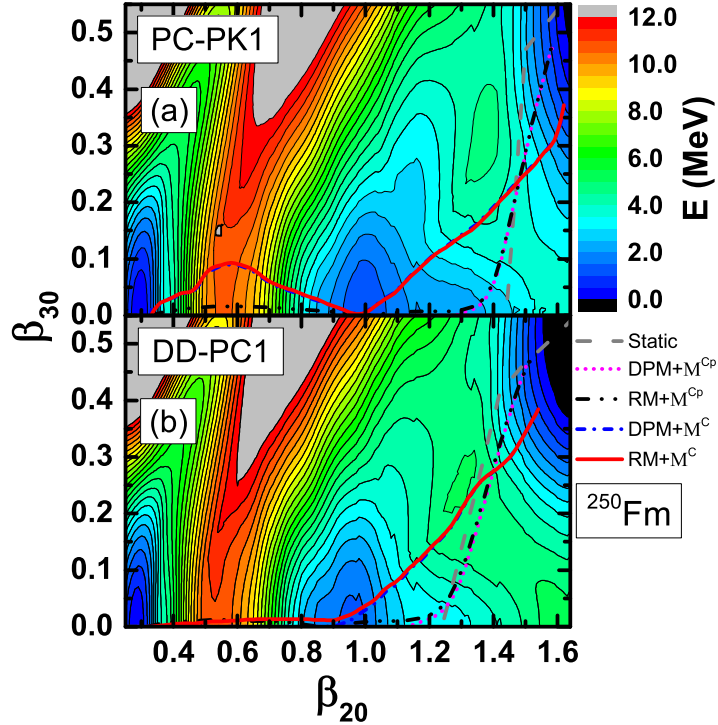}
\caption{(Color online)~\label{fig:Fm250_FT_path}%
Dynamic paths for spontaneous fission of $^{250}$Fm in the $(\beta_{20},\beta_{30})$
collective space with the perturbative and non-perturbative cranking 
inertia tensors. Both the dynamic-programming method and the Ritz method have been
used to minimize the fission action integral.
The static path (dashed curve) is also plotted for comparison. 
The results are obtained with the functionals PC-PK1 (upper panel) and DD-PC1
(lower panel), and axial symmetry is assumed.
}
\end{figure}

\begin{table}
%\vspace{-3mm}
%\footnotesize
%\begin{ruledtabular}
\begin{tabular}{llcr}
%\begin{tabular*}{170mm}{@{\extracolsep{\fill}}ccccccccc}
\hline \hline
EDF & Path                           & S(L)     & $\log_{10}(T_{1/2}/{\rm yr})$ \\ \hline
PC-PK1     & DPM+$\mathcal{M}^{Cp}$         & $28.23$  & $-3.52$                \\
           & RM+$\mathcal{M}^{Cp}$          & $28.23$  & $-3.52$                \\
           & DPM+$\mathcal{M}^{C}$          & $33.01$  & $0.64$                 \\
           & RM+$\mathcal{M}^{C}$           & $32.97$  & $0.60$                 \\
DD-PC1     & DPM+$\mathcal{M}^{Cp}$         & $30.74$  & $-1.34$                \\
           & RM+$\mathcal{M}^{Cp}$          & $30.73$  & $-1.35$                \\
           & DPM+$\mathcal{M}^{C}$          & $35.67$  & $2.95$                 \\
           & RM+$\mathcal{M}^{C}$           & $35.60$  & $2.89$                 \\
\hline \hline           
\end{tabular}
%\end{ruledtabular}
\caption{\label{tab:Fm250_FT} %
Values for the action integral and SF half-lives of $^{250}$Fm that correspond to the 
paths displayed in Fig.~\ref{fig:Fm250_FT_path}. The results 
obtained with the functionals PC-PK1 and DD-PC1 correspond to the axially-symmetric 
calculation in the $(\beta_{20},\beta_{30})$ plane.
}
\end{table}

In the case of $^{250}$Fm both the quadrupole triaxial $\beta_{22}$ 
and octupole $\beta_{30}$ collective degrees of
freedom play an important role in the spontaneous fission process. However, the calculation of
the dynamic fission path in the full 3D collective space ($\beta_{20}$, $\beta_{22}$ and
$\beta_{30}$), because of the huge number of computations required, 
is presently beyond our computational capabilities. For this reason, in the first step 
we determine the path in the restricted 2D collective space 
($\beta_{20},\beta_{30}$). The results are shown in Fig.~\ref{fig:Fm250_FT_path}, where
we compare the fission paths calculated using the functionals PC-PK1 (upper panel), and DD-PC1
(lower panel). For both cases the perturbative approach to calculating
the collective inertia produces a path which is close to the static one (minimum 
energy path). The dynamic path determined within the nonperturbative cranking inertia is 
markedly different and, for PC-PK1, it even makes an excursion in the reflection-asymmetric 
region already in the vicinity of the inner barrier. 
The corresponding 
action integrals and fission half-lives are listed in Tab.~\ref{tab:Fm250_FT}.
We note that in both cases the path passes through the isomeric state ($\beta_{20}\approx 0.95$,
$\beta_{30}=0$, $\beta_{22}=0$) which, in fact, is the common point for the two deformation spaces
($\beta_{20},\beta_{30}$) and ($\beta_{20},\beta_{22}$). Hence, the isomeric state presents the most
reasonable choice for the matching point at which the two paths are combined. 
\begin{figure}
 \includegraphics[width=0.6\textwidth]{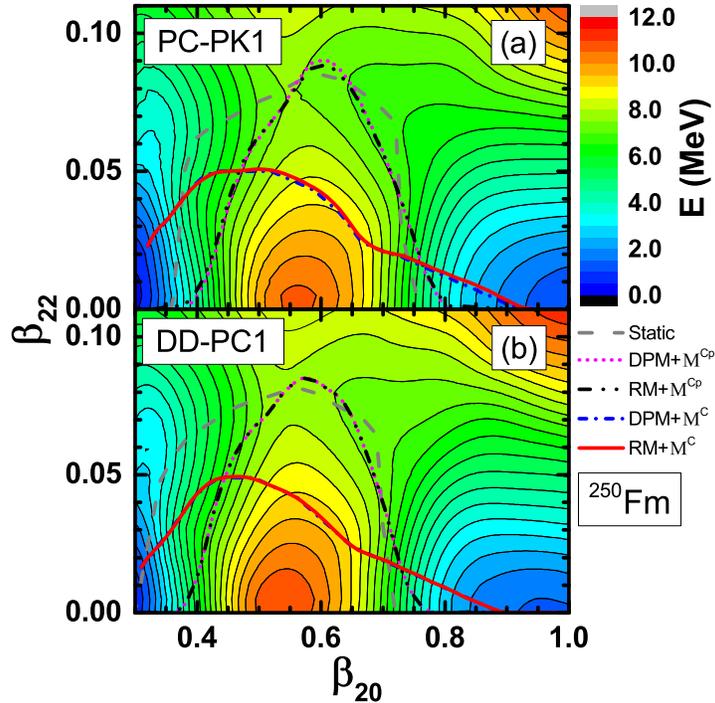}
\caption{(Color online)~\label{fig:Fm250_TF_path}%
Same as described in the caption to Fig.~\ref{fig:Fm250_FT_path} but for the 
paths in the $(\beta_{20},\beta_{22})$ plane. The shape is allowed
to be triaxial but reflection symmetric.
}
\end{figure}

\begin{table}
%\vspace{-3mm}
%\footnotesize
%\begin{ruledtabular}
\begin{tabular}{lllc}
\hline \hline
%\begin{tabular*}{170mm}{@{\extracolsep{\fill}}ccccccccc}
EDF & Symmetry & Path                           & S(L)      \\ \hline
PC-PK1     & {\sc axial}    & DPM+$\mathcal{M}^{Cp}$         & $20.61$                  \\
           &          & RM+$\mathcal{M}^{Cp}$          & $20.60$                  \\
           &          & DPM+$\mathcal{M}^{C}$          & $24.80$                   \\
           &          & RM+$\mathcal{M}^{C}$           & $24.46$                   \\
           & {\sc triaxial} & DPM+$\mathcal{M}^{Cp}$         & $19.57$                  \\
           &          & RM+$\mathcal{M}^{Cp}$          & $19.57$                  \\
           &          & DPM+$\mathcal{M}^{C}$          & $23.60$                   \\
           &          & RM+$\mathcal{M}^{C}$           & $23.54$                   \\
DD-PC1     & {\sc axial}    & DPM+$\mathcal{M}^{Cp}$         & $21.43$                   \\
           &          & RM+$\mathcal{M}^{Cp}$          & $21.42$                   \\
           &          & DPM+$\mathcal{M}^{C}$          & $25.65$                   \\
           &          & RM+$\mathcal{M}^{C}$           & $25.60$                   \\
           & {\sc triaxial} & DPM+$\mathcal{M}^{Cp}$         & $20.36$                  \\
           &          & RM+$\mathcal{M}^{Cp}$          & $20.35$                  \\
           &          & DPM+$\mathcal{M}^{C}$          & $24.50$                   \\
           &          & RM+$\mathcal{M}^{C}$           & $24.44$                   \\
\hline \hline           
\end{tabular}
%\end{ruledtabular}
\caption{\label{tab:Fm250_TF} %
Values of the action integral   
for different fission paths of $^{250}$Fm, connecting the inner turning point and the isomeric state. The label {\sc axial} denotes paths 
determined in the $(\beta_{20},\beta_{30})$ collective space, while 
the paths denoted with {\sc triaxial} correspond to the collective space $(\beta_{20},\beta_{22})$.}
\end{table}

As shown in the lower panel of Fig.~\ref{fig:Fm250_DDPC1_etot} the effects of triaxiality 
cannot be neglected in the region around the inner fission barrier. To take this degree of freedom 
into account we calculate the fission path in the $(\beta_{20},\beta_{22})$ collective space,
connecting the mean-field ground state and the isomeric state. The results are displayed in 
Fig.~\ref{fig:Fm250_TF_path}. Also in this case using the perturbative cranking 
collective inertia produces a path similar to the static one, whereas including the
nonperturbative cranking inertia modifies the path considerably. 
Although in the nonperturbative approach the paths do not reach that far in the triaxial region 
as the static and perturbative cranking dynamic paths, triaxial effects are obviously 
important for a realistic description of the spontaneous fission process of this isotope.
In Tab.~\ref{tab:Fm250_TF} we list the values of the action integral calculated
along the paths connecting the mean-field ground state and the isomeric state. 
The paths labeled with {\sc axial} and {\sc triaxial} are determined in the 
$(\beta_{20},\beta_{30})$ (cf. Fig.~\ref{fig:Fm250_FT_path}) and
$(\beta_{20},\beta_{22})$ (cf. Fig.~ \ref{fig:Fm250_TF_path}) collective space, respectively. 
One notices that, for both functionals, the inclusion of the triaxial degree of freedom 
reduces the value of the action integral.
For larger quadrupole $\beta_{20}$ deformations triaxial effects are less important, 
whereas the octupole degree of freedom plays a critical role in this region, 
as shown in Fig.~\ref{fig:Fm250_FT_path}.

Finally we combine the two segments: triaxial and reflection-symmetric from the inner 
turning point to the isomeric minimum at $\beta_{20}\approx 0.95$, and axial and 
reflection-asymmetric from the isomer to the outer turning point, to construct 
the entire dynamic fission path of $^{250}$Fm. The resulting 
action integrals and fission life-times are listed in Tab.~\ref{tab:action}.
For both functionals the action integrals calculated with the nonperturbative 
cranking collective inertia are considerably larger ($\approx 5$ units) than those 
obtained with the perturbative cranking inertia. Consequently, the nonperturbative
calculation predicts half-lives that are $\approx 4$ orders of magnitude longer in comparison
to the perturbative approach. One also notices that the functional DD-PC1 predicts 
larger action integrals and longer fission half-lives when compared to PC-PK1, both
for the perturbative and nonperturbative cranking inertia. This is consistent with the 
results obtained in the previous section for the symmetric fission of $^{264}$Fm. 
The effect of triaxiality on the asymmetric fission of $^{250}$Fm can be estimated by 
comparing the action integrals and fission life-times listed in Tab.~\ref{tab:Fm250_FT}
(axial and reflection-asymmetric from the inner to the outer turning points) 
and Tab.~\ref{tab:action} (triaxial and reflection symmetric up to the isomeric state, 
axial and reflection-asymmetric from the isomer to the outer turning point). We note that the 
inclusion of the triaxial degree of freedom lowers the value of the action integral 
and shortens the SF half-life of $^{250}$Fm, for both functionals and both approximations 
to the collective inertia. 

\begin{table}
%\vspace{-3mm}
%\footnotesize
%\begin{ruledtabular}
\begin{tabular}{llcr}
%\begin{tabular*}{170mm}{@{\extracolsep{\fill}}ccccccccc}
\hline \hline
EDF & Path                           & S(L)     & $\log_{10}(T_{1/2}/{\rm yr})$ \\ \hline
PC-PK1     & DPM+$\mathcal{M}^{Cp}$         & $27.19$  & $-4.42$                 \\
           & RM+$\mathcal{M}^{Cp}$          & $27.20$  & $-4.41$                 \\
            & DPM+$\mathcal{M}^{C}$          & $31.81$  & $-0.41$                 \\
          & RM+$\mathcal{M}^{C}$           & $32.05$  & $-0.20$                 \\
DD-PC1     & DPM+$\mathcal{M}^{Cp}$         & $29.67$  & $-2.27$                 \\
           & RM+$\mathcal{M}^{Cp}$          & $29.66$  & $-2.28$                 \\
         & DPM+$\mathcal{M}^{C}$          & $34.52$  & $1.95$                  \\
         & RM+$\mathcal{M}^{C}$           & $34.44$  & $1.88$                  \\
\hline \hline
\end{tabular}
%\end{ruledtabular}
\caption{\label{tab:action} %
Values for the action integral and SF half-lives of $^{250}$Fm that correspond to the 
triaxial and reflection-symmetric paths from the inner 
turning point to the isomeric minimum (cf. Fig.~ \ref{fig:Fm250_TF_path}), 
and axial and reflection-asymmetric from the isomer to the outer turning point(cf. Fig.~\ref{fig:Fm250_FT_path}).
}

\end{table}

%%%%%%%%%%%%%%%%%%%%%%%%%%%%%%%%%%%%%%%%%%%%%%%%%%
\section{\label{sec:summary}Summary and outlook}
%%%%%%%%%%%%%%%%%%%%%%%%%%%%%%%%%%%%%%%%%%%%%%%%%%

We have explored the dynamics of spontaneous fission of the nuclei $^{264}$Fm and
$^{250}$Fm in a theoretical framework based on relativistic energy density 
functionals. Deformation energy surfaces, collective potentials, and perturbative and nonperturbative 
ATDHFB cranking collective inertia tensors have been 
calculated with the multidimensionally-constrained relativistic Hartree-Bogoliubov (MDC-RHB) model, 
using the energy density functionals PC-PK1 and DD-PC1, and 
pairing correlations taken into account by a separable pairing force of finite range. 
Both the static (minimum energy) and dynamic (least action) 
fission paths, as well as the corresponding SF half-lives have been analyzed. 
For both nuclei considered in this study triaxial deformations lower the 
inner barrier by about $3$ MeV along the static fission path. The $^{264}$Fm isotope
undergoes symmetric fission into two $^{132}$Sn nuclei. Hence, this
process can be described in the 2D collective space spanned by the deformation coordinates 
$(\beta_{20},\beta_{22})$. The description of the asymmetric spontaneous fission of $^{250}$Fm, 
on the other hand, necessitates the inclusion of the octupole (reflection-asymmetric) 
degree of freedom $\beta_{30}$ and, in principle, calculations should be carried out in the full 
3D collective space spanned by the deformation coordinates $(\beta_{20},\beta_{22},\beta_{30})$.
In our case this is computationally too demanding and, therefore, the dynamic fission path of 
$^{250}$Fm is constructed from two segments: i) the path that connects the inner turning point
and the isomeric state is calculated by minimizing the fission action integral in the
$(\beta_{20},\beta_{22})$ plane, and ii) the path between the isomeric
state and the outer turning point is determined in the $(\beta_{20},\beta_{30})$ plane. 
The collective inertia tensors are calculated using the ATDHFB method both in the 
perturbative and nonperturbative cranking approximations.

Our study has confirmed previous results related to the perturbative approach to modelling the
ATDHFB collective inertia tensor. The perturbative treatment underestimates the effects of 
structural changes at the level crossing at which the nucleus changes its microscopic configuration 
diabatically, and the resulting collective inertia $\mathcal{M}^{Cp}$ varies relatively smoothly 
in the $(\beta_{20},\beta_{22})$ and $(\beta_{20},\beta_{30})$ planes. In contrast to the 
featureless behaviour of $\mathcal{M}^{Cp}$, the nonperturbative collective 
inertia $\mathcal{M}^{C}$ is characterized by the occurrence of sharp peaks 
on the surface of collective coordinates, that can 
be related to single-particle level crossings near the Fermi surface, that is, to abrupt changes 
of occupied single-particle configurations. This leads to an enhancement of the effective inertia, 
increases the fission action integral, and the resulting half-lives are longer. Consistent results have 
been obtained using the relativistic energy density functionals PC-PK1 and DD-PC1.

In the case of asymmetric fission of $^{250}$Fm we have analyzed the effect of triaxiality 
in the region around the inner fission barrier. Several recent studies have pointed out that 
nonaxial shapes are also relevant for the description of 
outer fission barriers in the actinides~\cite{Lu2012_PRC85-011301R,Lu2014_PRC89-014323}, 
and this will present an interesting topic for future applications of the model and computing 
methods developed in this work. An even more important issue is the inclusion of the 
particle-number fluctuation degree of freedom and the analysis of its impact on SF half-lives. 
Several recent studies have shown that pairing correlations should be treated on the same 
footing as shape deformation degrees of freedom, and we have also initiated work along these 
lines. 

%%%%%%%%%%%%%%%%%%%%%%%%%%%%%%%%%%%%%%%%%%
\appendix
\section{\label{Appendix-DPM-RM}Dynamic-programming method and Ritz method}
%----
\begin{figure}[t!]
 \includegraphics[width=0.6\textwidth]{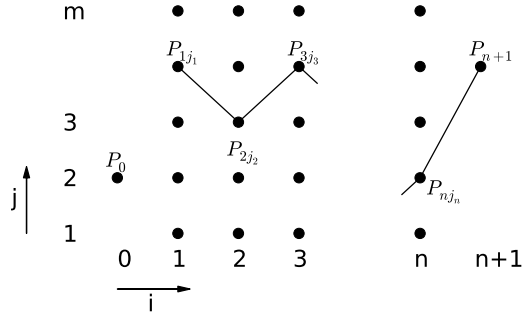}
\caption{\label{fig:DPM}%
Schematic diagram of the determination of the optimal trajectory using the dynamic-programming method.
}
\end{figure}
%-----

This appendix contains a brief description of the two methods used to find the minimum
action path in a two-dimensional collective space, for instance, on the $\beta_{20}\equiv q_0$, $\beta_{22}\equiv q_2$ plane.
Two locations are fixed on the path: the inner turning point $(q_0^{in},q_2^{in})$
and the outer turning point $(q_0^{out},q_2^{out})$.

The Ritz method (RM)~\cite{Baran1978_PLB76-8} is based on the variational procedure. 
The following ansatz is assumed for the path:
\begin{equation}
\label{eq:path-Ritz}
q_2(q_0) = \sum_{k=1}^N{a_k \sin{(k\pi x)}} + f(x),
\end{equation}
where the variable $x$ is defined:
\begin{equation}
x = \frac{q_2 - q_2^{in}}{q_2^{out}-q_2^{in}}, \quad 0 \le x \le 1.
\end{equation}
The ``boundary condition function" $f(x)$ ensures that the path passes through the inner and the
outer turning points, and in practical calculations a straight line can be used
\begin{equation}
f(x) = f_0 + (f_1 -f_0) x,
\end{equation}
with $ f_1 = f(1)$ and $f_0=f(0)$. The variational parameters $a_k$ are determined to
minimize the action integral calculated for the path defined in Eq.~(\ref{eq:path-Ritz}).
We note that the convergence of the Ritz method is rather fast, already with
$N=10$ stable results are obtained. 

The dynamic-programming method (DPM)~\cite{Baran1981_NPA361-83} uses 
an equidistant mesh in the plane of collective coordinates,
as shown in Fig.~\ref{fig:DPM}. The inner and outer turning points are denoted with
$P_0$ and $P_{n+1}$, respectively, and the path consists of straight line segments connecting the
mesh points $P_{ij}$. In the first step one determines the optimal paths connecting the 
inner turning point $P_0$ and each point in the second column of the mesh $P_{2j_2}$ 
($j_2=1,\dots,m$). This is achieved by comparing the action integrals for $m$ paths 
passing through separate points $P_{1j_1}$ in the first column of the mesh and, in this way, 
for each point in the second column $P_{2j_2}$ an optimal trajectory is obtained which 
minimizes the action integral. In the second step the optimal path from the inner turning
point $P_0$ to each point in the third column $P_{3j_3}$ ($j_2=1,\dots,m$) is determined, again by 
comparing only $m$ paths passing through various points in the second column $P_{2j_2}$.
This procedure is repeated until the outer turning point is reached, that is, the entire optimal path
is constructed. The main advantage of this method is that one has to calculate and compare
only $m\times n$ paths from the $m^n$ possible paths on the mesh. The following values for the 
mesh size have been used in the present analysis: 0.01 for the coordinate $\beta_{20}$, 
0.001 for $\beta_{22}$, and 0.002 for $\beta_{30}$. 

\bigskip
%---------------------------------------------------------
\acknowledgements
This work has been supported in part by the NEWFELPRO project
of the Ministry of Science, Croatia, co-financed through the Marie
Curie FP7-PEOPLE-2011-COFUND program, and by
the Croatian Science Foundation -- project ``Structure and Dynamics
of Exotic Femtosystems" (IP-2014-09-9159).
We thank Jhilam Sadhukhan for very helpful discussions.
Calculations have been performed in part at  
the High-performance Computing Cluster of SKLTP/ITP-CAS.

%\bibliographystyle{apsrev4-1}
%\bibliography{Nuclear-Fission}


\end{document}